\documentclass[letter,twocolumn]{jpsj3}

\usepackage[dvipdfmx]{graphicx}
\usepackage[dvipdfmx]{color}
\usepackage{amssymb}
\usepackage{bm}
\usepackage{ulem}

\title{
Kink in Stoner Factor as a Signature of Changing Magnetic Fluctuations in the Organic Conductor $\lambda$-(BETS)$_2$GaCl$_4$
}

\author{
Hirohito Aizawa\thanks{h-aizawa@uitec.ac.jp, h.aizawa.phys@gmail.com}
}

\inst{
Division of Basic Human Resources Development, 
The Polytechnic University of Japan, 
Kodaira, Tokyo 187-0035, Japan \\
}

\abst{
In this study, we theoretically investigated the magnetic properties of the quasi-two-dimensional organic conductor $\lambda$-(BETS)$_2$GaCl$_4$ using a multi-band Hubbard model and the two-particle self-consistent method. 
We employed a four-band model in which each BETS molecule is considered as a site and a two-band model that considers each BETS dimer as a site. 
Our results for the temperature dependence of the Stoner factor reveal a kink around $T_\mathrm{kink} \approx 5~\mathrm{meV}$, which indicates a change in the dominant magnetic fluctuations. 
A broad structure indicating smeared antiferromagnetic~(AFM) fluctuations was observed above $T_\mathrm{kink}$, whereas the spin susceptibility peaked at a wavevector corresponding to spin-density-wave~(SDW)-like fluctuations below $T_\mathrm{kink}$. 
The kink disappears as the intra-dimer transfer integral increased and the AFM fluctuations were enhanced. 
Our findings are consistent with those of previous experimental observations, which have also reported a change in magnetic properties from AFM to SDW-like fluctuations upon cooling. 
}

\begin{document}
\maketitle

The organic conductor $\lambda$-(BETS)$_2$GaCl$_4$, where BETS stands for bis(ethylenedithio)tetraselenafulvalene, exhibits metallic behavior at ambient pressure and undergoes a superconducting transition at approximately 5.5~K upon cooling~\cite{H_Kobayashi1993, A_Kobayashi1993}. 
Various experimental studies have reported the presence of anisotropic superconductivity~(SC) in this material~\cite{Tanatar1999, Yasuzuka2014, Imajo2016, Imajo2019, Kobayashi2020_d-wave_NMR, Dita2021}, which has also been further supported by a theoretical investigation~\cite{Aizawa2018}. 
Furthermore, several studies have suggested that the Fulde–Ferrell–Larkin–Ovchinnikov superconducting state arises near the upper critical field~\cite{Tanatar2002, Coniglio2011, Uji2015, Imajo2021}. 

Recently, the magnetic properties of $\lambda$-(BETS)$_{2}$GaCl$_{4}$ have been actively investigated in studies on nuclear magnetic resonance~(NMR). 
These studies have employed various nuclei, including $^{77}$Se~\cite{Takagi2003_77Se_NMR}, $^{1}$H~\cite{Takagi2003_1H_NMR}, $^{13}$C~\cite{Kobayashi2017}, and $^{69, 71}$Ga~\cite{Kobayashi2020_no_increase_in_charge-fluctuation_NMR} for the compound under ambient pressure. 
Several studies have reported the development of antiferromagnetic~(AFM) fluctuations in the temperature range of several tens of kelvin to slightly above 100~K~\cite{Takagi2003_77Se_NMR, Takagi2003_1H_NMR, Kobayashi2017, Kobayashi2020_no_increase_in_charge-fluctuation_NMR, Kobayashi2020_chem-press_NMR}. 
Additionally, recent experiments have reported magnetic behavior associated with spin-density-wave (SDW) phenomena at low temperatures. 
Specifically, magnetic fluctuations originating  from the Fermi surface nesting~\cite{Kobayashi2017} or SDW fluctuations~\cite{Kobayashi2020_no_increase_in_charge-fluctuation_NMR, Sawada2021} develop at low temperatures. 
In $\lambda$-(BETS)$_{2}$GaBr$_{0.75}$Cl$_{3.25}$, the SDW order emerges at lower temperatures, as observed using $^{13}$C NMR~\cite{Kobayashi2020_chem-press_NMR}. 
A theoretical study has suggested that the Mott insulating state is accompanied by an AFM order in $\lambda$-(BETS)$_{2}$GaBr$_{z}$Cl$_{4-z}$~\cite{Seo1997}. 
This has been demonstrated using a Hubbard model in which each BETS molecule is considered  a site and a localized spin system in which a BETS dimer is regarded as a site. 
Clarifying the magnetic properties is essential not only to understand the possible insulating phase itself but also to elucidate the mechanism of the adjacent SC. 

\begin{figure}[!htb]
\centering
\includegraphics[width=8.0cm]{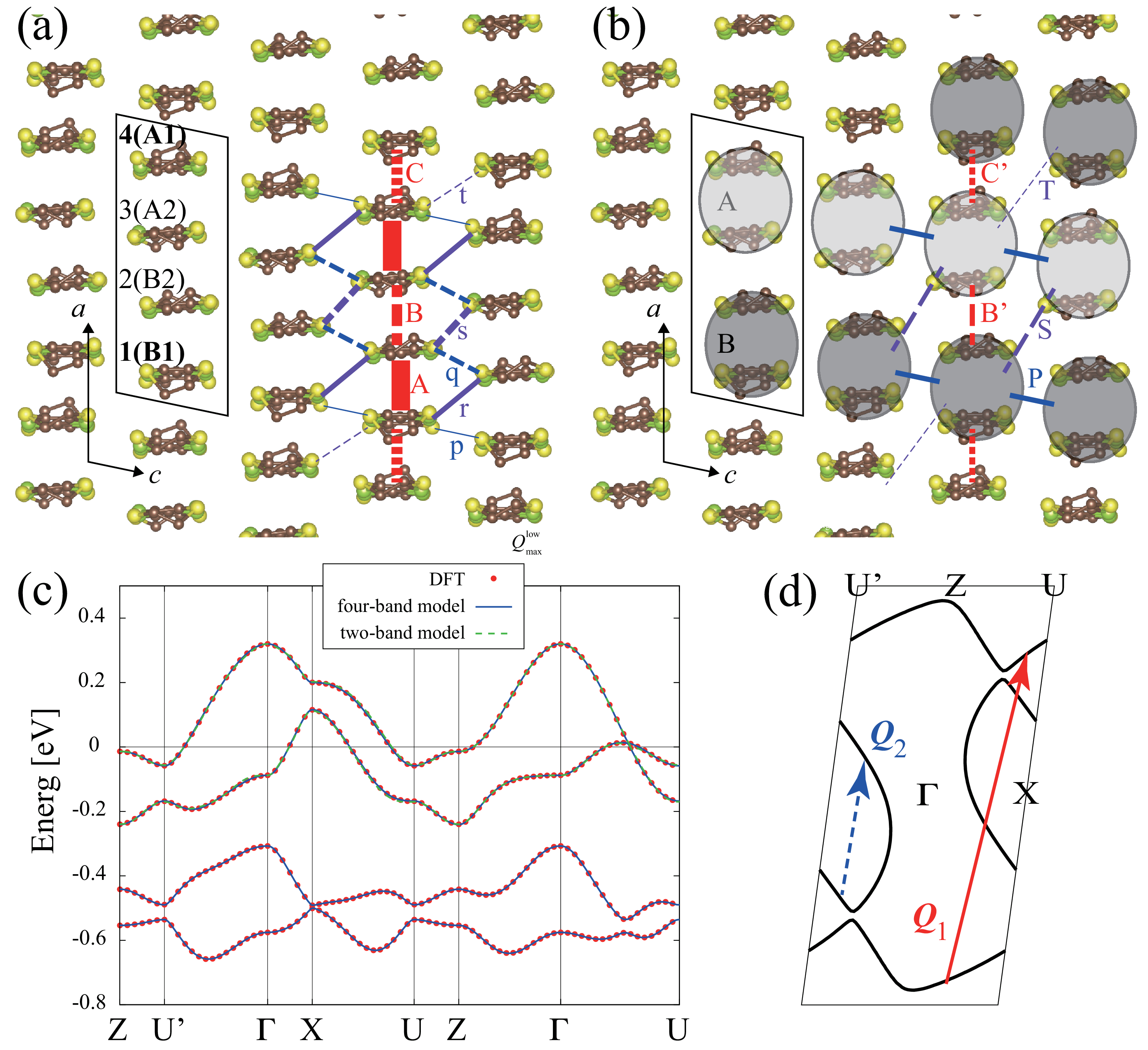}
\caption{
(Color online) Crystal and electronic structure of $\lambda$-(BETS)$_2$GaCl$_4$. (a) Four-band model in which each BETS molecule is treated as a site, and (b) two-band model in which a BETS dimer is treated as a site. 
(c) Band structure, where the red dotted curves represent the DFT band structure, and the blue solid (green dashed) curves correspond to the four-band (two-band) model~\cite{Aizawa2018}. 
(d) Fermi surface obtained from the four-band model. The red solid (blue dashed) arrow $\bm{Q}_1$ ($\bm{Q}_2$) represents the characteristic wavenumber vector in the first Brillouin zone. These wavenumber vectors correspond to the dominant magnetic fluctuations discussed later~(see Fig.~\ref{fig3}). 
}
\label{fig1}
\end{figure}

As shown in Fig.~\ref{fig1}(a), the crystal structure of $\lambda$-(BETS)$_2$GaCl$_4$ comprises a conducting BETS donor molecular layer and a GaCl$_4^{-1}$ anion layer. 
This arrangement forms a quasi-two-dimensional electronic system within the BETS layer. 
Given the composition ratio and molecular valence, the BETS donor molecules are 1/4-filled in terms of holes, which corresponds to being 3/4-filled in terms of electrons. 
We refer to the model in which each BETS molecule is treated as a site as the four-band model. This is shown in Fig.~\ref{fig1}(a). 
Conversely, in the limit of strong dimerization of the BETS molecules, the configuration in which each BETS dimer is treated as a site is referred to as the two-band model. This is shown in Fig.~\ref{fig1}(b). 
For comparison with the magnetic properties in the four-band model, the two-band model is justified by the strong dimerization in which the intra-dimer transfer integral~$t_\mathrm{A} \left( \approx 230~\mathrm{meV} \right)$ is significantly larger than inter-dimer transfer integrals~$\left( \approx 130~\mathrm{meV} \right)$ as confirmed by the density functional theory~(DFT) calculations provided in our previous work~\cite{Aizawa2018}. 
This two-band model is half-filled in terms of electrons per dimer. 
As shown in Fig.~\ref{fig1}(c) and Fig.~\ref{fig1}(d), the Fermi surface obtained from the four-band model consists of a quasi-one-dimensional open Fermi surface and a cylindrical closed Fermi surface around the X point. 
A microscopic theory for the SC gap symmetry suggests a $d$-wave-like gap mediated by spin fluctuations~\cite{Aizawa2018}. 
Thus, elucidating the magnetic properties is crucial to clarify the mechanism of SC in $\lambda$-(BETS)$_{2}$GaCl$_{4}$. 

In this study, we present the temperature dependence of the Stoner factor for spin susceptibility, which we calculated using several four-band models and the two-band model. 
We then analyze the spin susceptibility of selected parameter sets and discuss the resulting magnetic structures in relation to experimental observations.  

Here, we introduce the Hubbard Hamiltonian $H$ based on the multi-site tight-binding model represented as 
\begin{eqnarray}
 H&=&\sum_{\left< i \alpha: j \beta \right>, \sigma}
  \left\{ t_{i \alpha: j \beta} 
   c_{i \alpha \sigma}^{\dagger} c_{j \beta \sigma} + {\rm H. c.} 
  \right\} 
\nonumber \\
  &+&
  \Delta E \sum_{i, \alpha=2, 3} n_{i \alpha}
  +\sum_{i, \alpha} U
  n_{i \alpha \uparrow} n_{i \alpha \downarrow}, 
  \label{Hij}
\end{eqnarray} 
where $i$ and $j$ are unit-cell indices, $\alpha$ and $\beta$ specify sites in the unit cell, $c_{i \alpha \sigma}^{\dagger}$ ($c_{i \alpha \sigma}$) is a creation (annihilation) operator for spin $\sigma$ at site $\alpha$ in unit cell $i$, $t_{i \alpha: j \beta}$ is the transfer integral between site $(i, \alpha)$ and site $(j, \beta)$, and $\left< i \alpha: j \beta \right>$ represents the summation over bonds that corresponds to the transfer integrals. 
Here, $\Delta E$ represents the energy difference between the BETS molecules 1(4) and 2(3) as shown in Fig.~\ref{fig1}(a). 
Note that $\Delta E=0$ in the two-band model. 
$U$ is the on-site interaction, and $n_{i \alpha \sigma}$ is the number operator for electrons with spin $\sigma$ on site $\alpha$ in unit cell $i$. 
As shown in Fig.~\ref{fig1}(c), the band is 3/4-filled (half-filled) in the electron representation in the four-band (two-band) model. 
The transfer integrals were derived from the DFT calculations as described in our previous work~\cite{Aizawa2018}. 

To account for the effects of electron correlation, we employed the two-particle self-consistent (TPSC)~\cite{Vilk1997} method in our study of the multi-band Hubbard model. 
The TPSC method has been applied to single-band~\cite{Vilk1997,Otsuki2012}, multi-band~\cite{Arya2015, Ogura2015, Aizawa2015, Aizawa2018PRB, Aizawa2018JPCS}, and multi-orbital systems~\cite{Miyahara2013}. 
In the TPSC method, the spin susceptibility matrix 
$\hat{\chi}^\mathrm{sp}\left( \bm{q}, i \omega_{m} \right)$ 
and the charge susceptibility matrix 
$\hat{\chi}^\mathrm{ch}\left( \bm{q}, i \omega_{m} \right)$ 
are given by 
\begin{eqnarray}
\hat{\chi}^\mathrm{sp}\left( \bm{q}, i \omega_{m} \right) 
 &=&\left[ \hat{I}-\hat{\chi}^{0}\left( \bm{q}, i \omega_{m} \right) \hat{U}^\mathrm{sp} \right]^{-1} \hat{\chi}^{0}\left( \bm{q}, i \omega_{m} \right),
 \label{chisp-tpsc} 
 \\
 \hat{\chi}^\mathrm{ch}\left( \bm{q}, i \omega_{m} \right)
 &=&\left[ \hat{I}+\hat{\chi}^{0}\left( \bm{q}, i \omega_{m} \right) \hat{U}^\mathrm{ch} \right]^{-1} \hat{\chi}^{0}\left( \bm{q}, i \omega_{m} \right). 
\end{eqnarray}
Given that the spin susceptibility and the Stoner factor reach their maximum values at $i\omega_{m}=0$, we simplify the notation by omitting the Matsubara frequency, representing $\left( \bm{q},  i \omega_{m} \right)$ simply as $\left( \bm{q} \right)$ hereafter. 
These local vertices are determined self-consistently by two sum rules for the local moment and an ansatz for double occupancy~\cite{Vilk1997}. 
We define the spin susceptibility $\chi_\mathrm{sp}$ as the largest eigenvalue of $\hat{\chi}^\mathrm{sp}$, whereas the Stoner factor $\alpha_\mathrm{S}\left( \bm{Q}_\mathrm{max} \right)$ is obtained as that of $\hat{\chi}^\mathrm{0}\left( \bm{Q}_\mathrm{max} \right) \hat{U}_\mathrm{sp}$, where $\bm{Q}_\mathrm{max}$ is the nesting vector. 
The magnetic transition temperature $T_\mathrm{c}$ is defined as the temperature at which $\alpha_\mathrm{S}\left( \bm{Q}_\mathrm{max} \right)$ reaches unity. 
A true magnetic transition does not occur in a purely two-dimensional system because the TPSC method satisfies the Mermin-Wagner theorem. 
Therefore, the temperature at which $\alpha_\mathrm{S}\left( \bm{Q}_\mathrm{max} \right)$ approaches unity is considered as the magnetic critical temperature in the actual three-dimensional system. 
Although the qualitative kink structure reflects the temperature dependence of the bare susceptibility ($\hat{\chi}^0$), the interaction which includes the electron correlation ($\hat{U}^\mathrm{sp}$) obtained from the TPSC approach is crucial for quantitative results. This is necessary to capture the quantitative temperature scale~\cite{Aizawa2015, Aizawa2018PRB, Aizawa2018JPCS} of the $T_\mathrm{kink}$ and to consider the suppression of the magnetic correlation due to enhanced double occupancy. 
We used a system size of $96 \times 96$ $k$-meshes and $16384$ Matsubara frequencies. 
In the four-band Hubbard model, we set the on-site interaction $U$ to $1~\mathrm{eV}$, which is approximately the same as the bandwidth of the four-band model. 
The intra-dimer transfer integral $t_\mathrm{A}$ was a variable parameter. 
In the two-band Hubbard model, $U$ was set to $0.8~\mathrm{eV}$, which was slightly larger than the bandwidth. 
These $U$ values are supported by previous theoretical studies that evaluated the Coulomb interactions in several organic conductors~\cite{Nakamura2009, Nakamura2016, Misawa2020, Kato2025} using the constrained random phase approximation~\cite{Aryasetiawan2004, Imada2010}. 

\begin{figure}[!htb]
\centering
\includegraphics[width=8.0cm]{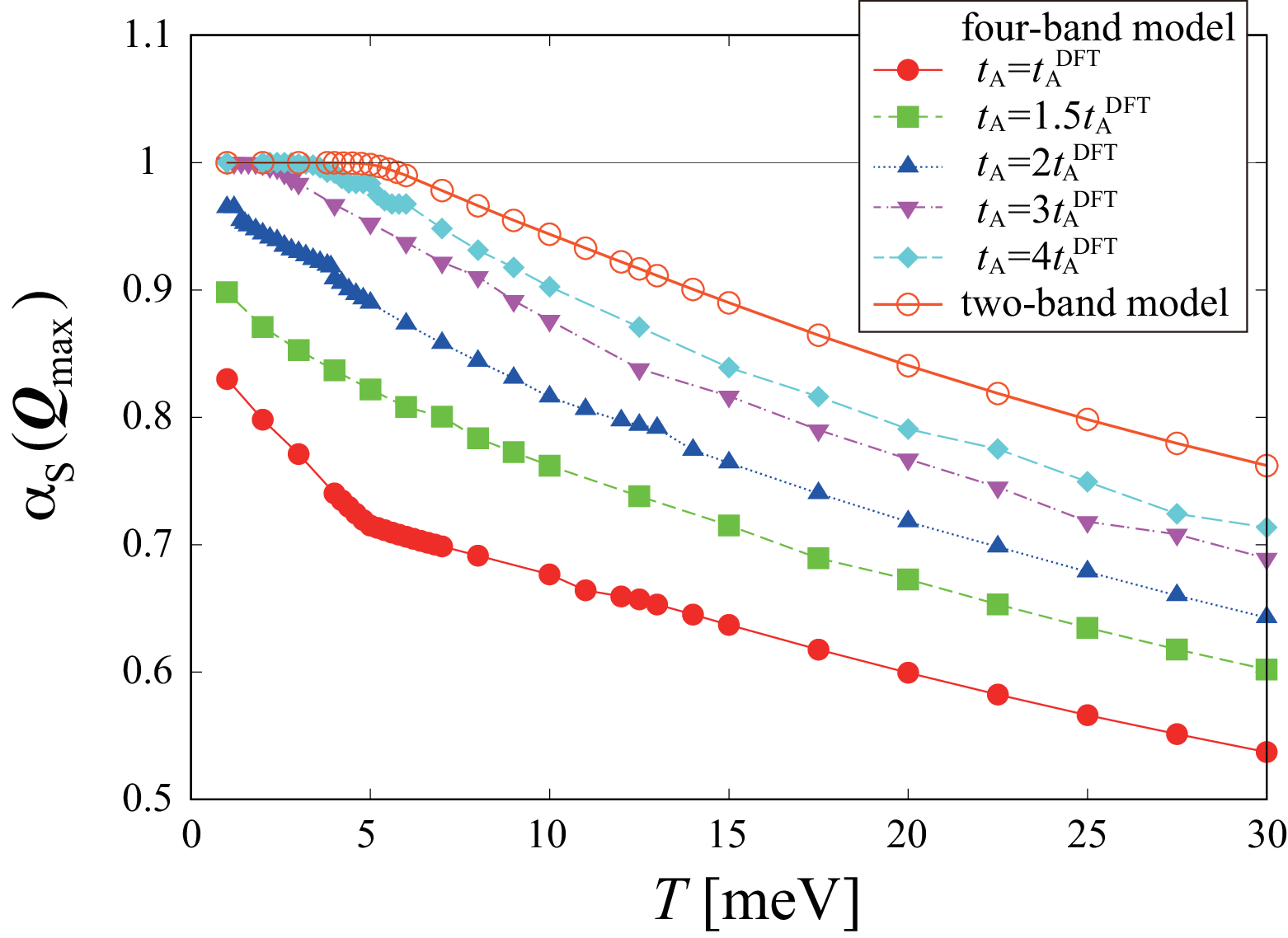}
\caption{
(Color online) Temperature dependence of the Stoner factor $\alpha_\mathrm{S}\left( \bm{Q}_\mathrm{max} \right)$ for various transfer integrals in the four-band and two-band models, with symbols and curves as indicated in the legend. 
}
\label{fig2}
\end{figure}

The results are shown in Fig.~\ref{fig2}, where the temperature dependence of the Stoner factor $\alpha_\mathrm{S}\left( \bm{Q}_\mathrm{max} \right)$ is presented. 
In the four-band model in which the intra-dimer transfer integral $t^\mathrm{DFT}_\mathrm{A}$ is obtained from DFT calculations,  the results show that a kink appears around $T_\mathrm{kink} \approx 5~\mathrm{meV}$. 
The slope of the temperature dependence of the $\alpha_\mathrm{S}\left( \bm{Q}_\mathrm{max} \right)$ changes across $T_\mathrm{kink}$. 
Lowering the temperature increases $\alpha_\mathrm{S}\left( \bm{Q}_\mathrm{max} \right)$; however, $\alpha_\mathrm{S}\left( \bm{Q}_\mathrm{max} \right)$ does not reach unity. 
Thus, within the scheme and parameter set in the present study, magnetic fluctuations develop upon cooling and the magnetic properties change across $T_\mathrm{kink}$, although no magnetic transition occurs in the four-band model. 
These magnetic properties seem to be consistent with the actual magnetic behavior of $\lambda$-(BETS)$_{2}$GaCl$_{4}$ at the ambient pressure. 

Increasing the intra-dimer transfer integral $t_\mathrm{A}$, which corresponds to enhanced dimerization, leads to the disappearance of the kink and a monotonous increase in the $\alpha_\mathrm{S}\left( \bm{Q}_\mathrm{max} \right)$ across the entire temperature regime. 
In the model in which $t_\mathrm{A}$ is three times larger than the original value $t_\mathrm{A}^\mathrm{DFT}$, 3$t_\mathrm{A}^\mathrm{DFT}$, the saturation of the $\alpha_\mathrm{S}\left( \bm{Q}_\mathrm{max} \right)$ was evident in the low-temperature regime. 
Thus, an increase in the dimerization led to the appearance of a magnetic transition. 
In the two-band model that corresponds to the dimer limit, $\alpha_\mathrm{S}\left( \bm{Q}_\mathrm{max} \right)$ increases with decreasing temperature and saturates to unity around $T_\mathrm{c} \approx 5~\mathrm{meV}$. 
Thus, a magnetic transition appeared around $T_\mathrm{c} \approx 5~\mathrm{meV}$ in the two-band model for the parameter set that we used. 

\begin{figure}[!htb]
\centering
\includegraphics[width=8.0cm]{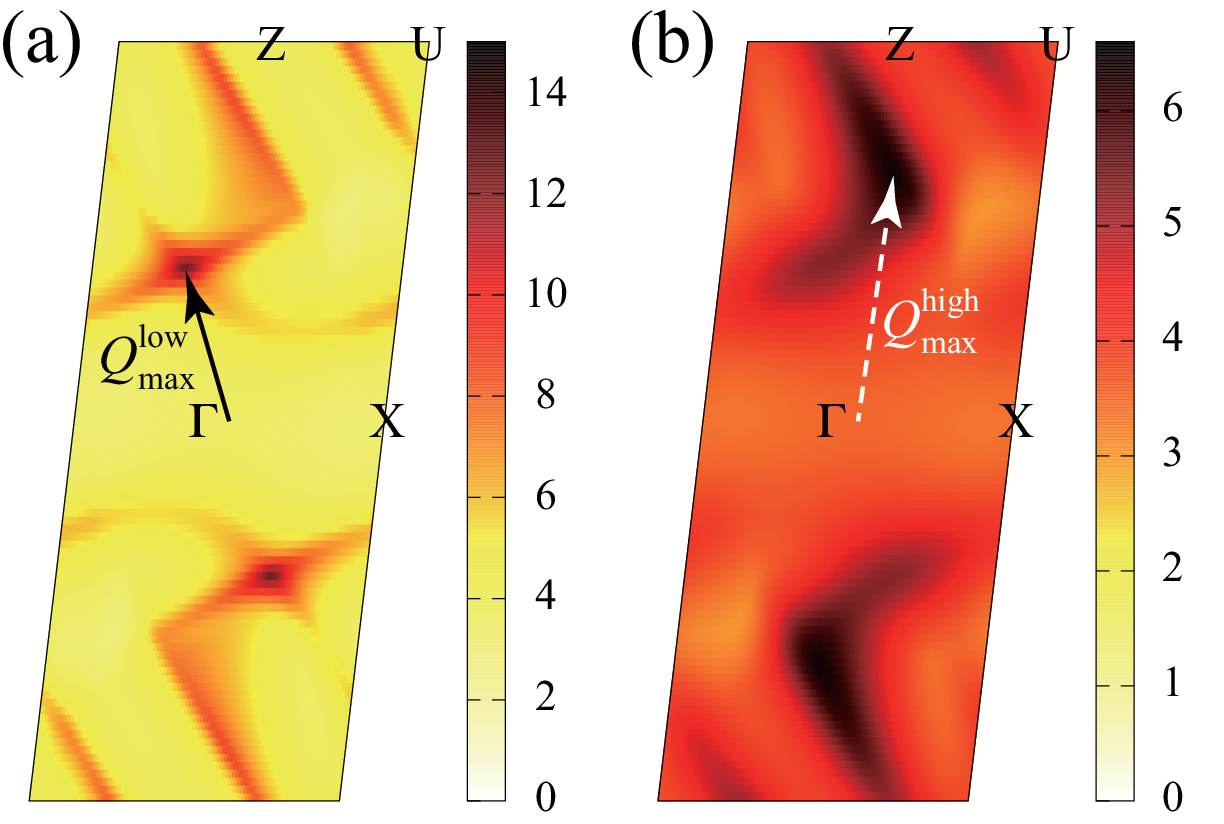}
\caption{
(Color online) The spin susceptibility $\chi_\mathrm{sp}\left( \bm{q} \right)$ of the four-band model with $t_\mathrm{A}=t^\mathrm{DFT}_\mathrm{A}$ at (a) a temperature below $T_\mathrm{kink}$ ($T=2~\mathrm{meV}$) and (b) a temperature above $T_\mathrm{kink}$ ($T=10~\mathrm{meV}$), where $T_\mathrm{kink} \approx 5~\mathrm{meV}$. 
}
\label{fig3}
\end{figure}

Fig.~\ref{fig3}(a) shows the spin susceptibility of the four-band model with the original intra-dimer transfer integral $t_\mathrm{A}^\mathrm{DFT}$ at $T=2~\mathrm{meV}$, which is lower than $T_\mathrm{kink}$. 
The wavenumber giving the maximum spin susceptibility is $\bm{Q}_\mathrm{max} \approx \left( -\frac{3}{8}\pi, \frac{3}{8}\pi \right)$. 
We define this wavenumber vector as $\bm{Q}_\mathrm{max}^\mathrm{low}$ which  corresponds to $\bm{Q}_1$ in Fig.~\ref{fig1}(d). 
In real space, the $\bm{Q}_\mathrm{max}^\mathrm{low}$ corresponds to long-periodic magnetic fluctuations such as SDW fluctuations along both the crystal $c$- and $a$-axes. 
Fig.~\ref{fig3}(b) shows the spin susceptibility at $T=10~\mathrm{meV}$ which is higher than $T_\mathrm{kink}$. 
The wavenumber of the maximum spin susceptibility is obtained as $\bm{Q}_\mathrm{max} \approx \left( 0, \frac{2}{3}\pi \right)$. 
We define this wavenumber vector as $\bm{Q}_\mathrm{max}^\mathrm{high}$ which  corresponds to $\bm{Q}_2$ in Fig.~\ref{fig1}(d). 
In addition, the spin susceptibility exhibits a broad structure in wavenumber space, which extends from $\bm{Q}_\mathrm{max}^\mathrm{high}$ to a wavenumber slightly shifted from $\bm{q} = (0,\pi)$. 
We suggest that the system exhibits short-range periodic magnetic fluctuations reminiscent of the AFM fluctuations. 
The spin modulation between unit cells is inferred to be staggered along the $c$-axis, while remaining uniform along the $a$-axis. 

\begin{figure}[!htb]
\centering
\includegraphics[width=8.0cm]{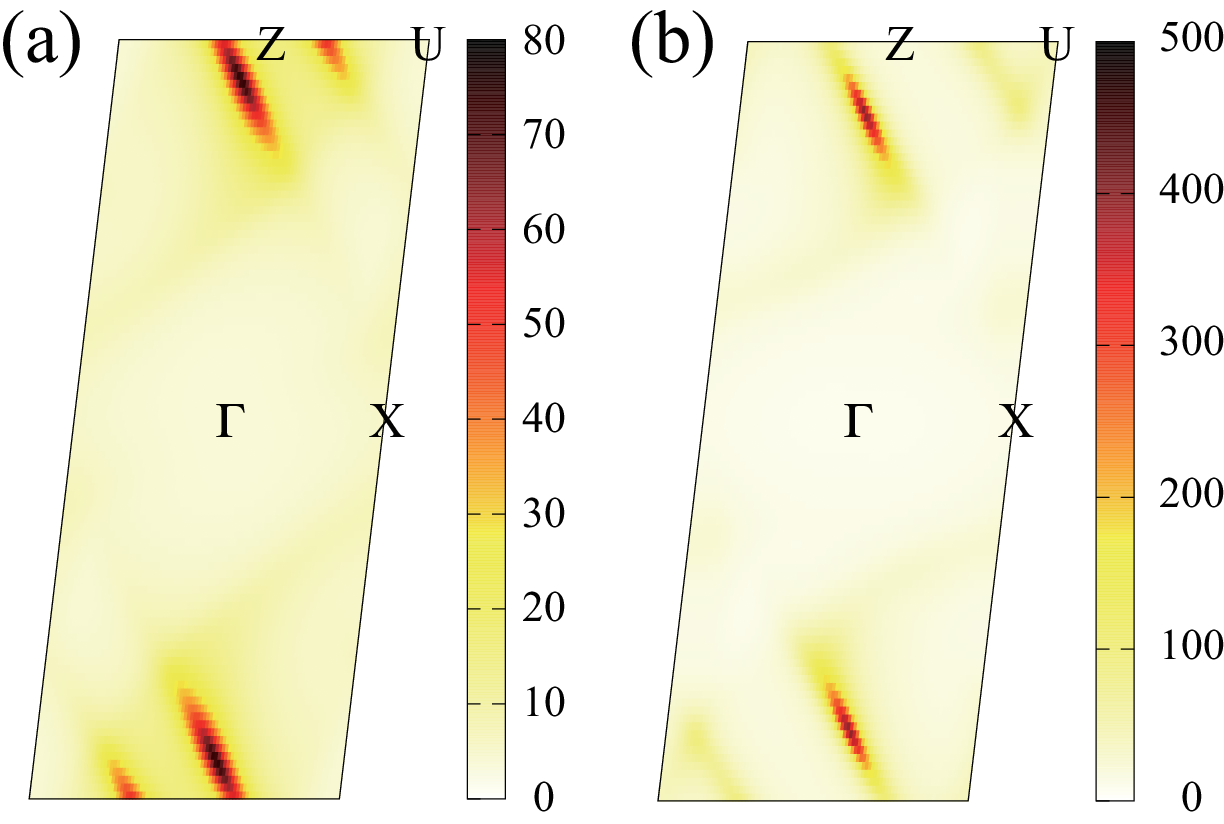}
\caption{
(Color online) The spin susceptibility $\chi_\mathrm{sp}\left( \bm{q} \right)$ at $T=6~\mathrm{meV}$ for (a) the four-band model with $t_\mathrm{A}=4 t_\mathrm{A}^\mathrm{DFT}$, and (b) the two-band model. 
}
\label{fig4}
\end{figure}

Fig.~\ref{fig4}(a) shows the spin susceptibility of the four-band model with $t_\mathrm{A}$ set to $4 t_\mathrm{A}^\mathrm{DFT}$ at $T=6~\mathrm{meV}$, which corresponds to the temperature at which $\alpha_\mathrm{S}\left( \bm{Q}_\mathrm{max} \right)$ nearly reaches unity as shown in Fig.~\ref{fig2}. 
As shown in Fig.~\ref{fig4}(a), the peak in the spin susceptibility near $\bm{Q}_\mathrm{max}^\mathrm{low}$, as shown in Fig.~\ref{fig3}(a) was not observed. 
Instead, the maximum spin susceptibility occurs at $\bm{Q}_\mathrm{max} \approx \left( 0, \pi \right)$. 
In real space, spin fluctuations suggest short-range periodic magnetic correlations that are modulated along $c$-axis. 
Thus, the stronger dimerization enhances spin fluctuations that are staggered along the $c$-axis but uniform along the $a$-axis, reminiscent of the AFM fluctuations in the four-band model with $4 t_\mathrm{A}^\mathrm{DFT}$. 

Fig.~\ref{fig4}(b) shows the spin susceptibility in the two-band model, which corresponds to the limit at which $t_\mathrm{A}$ approaches infinity in the four-band model. 
The result is shown at $T=6~\mathrm{meV}$, where $\alpha_\mathrm{S}\left( \bm{Q}_\mathrm{max} \right)$ begins to saturate at unity. 
This spin susceptibility closely resembles that of the four-band model with a large $t_\mathrm{A}$, as shown in Fig.~\ref{fig4}(a), where the maximum occurs at approximately $\bm{Q}_\mathrm{max} \approx \left( 0, \pi \right)$. 
In real space, this indicates the AFM fluctuations in the two-band model. 

For the purpose of discussion, the NMR experiment on $\lambda$-(BETS)$_2$GaCl$_4$~\cite{Kobayashi2017} revealed a Curie-like increase in $1/T_{1}T$ above 55~K, suggesting the development of AFM fluctuations. 
Below 10~K, $1/T_{1}T$ was further enhanced, which indicates that magnetic fluctuations were induced by Fermi surface nesting. 
Another NMR study on $\lambda$-(BETS)$_2$GaBr$_{0.75}$Cl$_{3.25}$~\cite{Kobayashi2020_chem-press_NMR} reported that $1/T_{1}T$ exhibits a kink at 30~K, which increases again below 25~K and diverges at 13~K, indicating the onset of the SDW state. 

As shown in Fig.~\ref{fig2}, a comparison between the experimental results and our findings in the present work reveals a similar temperature dependence of the spin susceptibility, with a kink appearing around $T_\mathrm{kink}\approx 5~\mathrm{meV}$. 
As shown in Fig.~\ref{fig3}, the magnetic properties vary across $T_\mathrm{kink}$. 
Above $T_\mathrm{kink}$ as shown in Fig.~\ref{fig3}~(b), the spin susceptibility exhibits a broad structure extending from $\bm{Q}_\mathrm{max}^\mathrm{high}$ to approximately $\left( 0, \pi \right)$, reminiscent of the AFM fluctuations, whereas below $T_\mathrm{kink}$, the SDW-like fluctuations develop with $\bm{Q}_\mathrm{max}^\mathrm{low}$ as shown in Fig.~\ref{fig3}~(a). 
To further investigate the broad wavenumber structure shown in Fig.~\ref{fig3}~(b), we analyzed the effect of increasing the intra-dimer transfer integral in the four-band model. 
As shown in Fig.~\ref{fig4}~(a), this increase enhances the spin susceptibility around $\left( 0, \pi \right)$. 
In the two-band model, which corresponds to the dimer limit of the four-band model, the spin susceptibility exhibits a maximum around $\left( 0, \pi \right)$. 
This suggests that the broad wavenumber structure observed near $\bm{q} = \left( 0, \pi \right)$ above $T_\mathrm{kink}$ in the four-band model can be interpreted as ``smeared AFM fluctuations''. 

Our results are in agreement with those of studies on NMR~\cite{Kobayashi2017}, such as the dominant AFM fluctuations observed above $T \approx 50~\mathrm{K}$ and the development of SDW-like fluctuations at low temperatures. 
However, a difference was observed in the intermediate regime ($10~\mathrm{K} < T < 50~\mathrm{K}$), where magnetic fluctuations are suppressed. 
The simplicity of our model, in which we consider only the on-site interaction $U$, seems to account for this discrepancy, which is the suppression of the SDW-like fluctuations within the intermediate temperature regime. 
This difference suggests that the actual material involves factors absent from our calculation, namely, the magnetic competition induced by the semiconductor-metal crossover suggested by an experimental study~\cite{Kobayashi2017}, the effects of lattice degrees of freedom and long-range Coulomb interactions, and the effects of disorder. 
Investigating the interplay between these effects is important for future theoretical studies. 

In conclusion, we performed the TPSC calculations of spin susceptibility in four- and two-band Hubbard models of the quasi-two-dimensional organic conductor $\lambda$-(BETS)$_2$GaCl$_4$ using transfer integrals obtained from DFT calculations. 
The Stoner factor $\alpha_\mathrm{S}\left( \bm{Q}_\mathrm{max} \right)$ exhibits a kink at approximately $T_\mathrm{kink} \approx 5~\mathrm{meV}$, which indicates a change in both the magnitude and wavenumber of the maximum spin susceptibility. 

Below $T_\mathrm{kink}$, the spin susceptibility peaks at $\bm{Q}_\mathrm{max}^\mathrm{low}$, which corresponds to the SDW-like fluctuations along both the $c$- and $a$-axes. 
Above $T_\mathrm{kink}$, it shifts to $\bm{Q}_\mathrm{max}^\mathrm{high}$, with a broad structure around $\bm{Q}_\mathrm{max} \approx \left( 0, \pi \right)$, which resembles smeared AFM fluctuations along the $c$-axis. 
As the intra-dimer transfer integral $t_\mathrm{A}$ increases, the kink in the temperature dependence of the Stoner factor disappears. 
In the two-band model, the Stoner factor is larger and saturates to unity around $T_\mathrm{kink}$. 
In both models with large or diverging $t_\mathrm{A}$, the maximum spin susceptibility occurs at $\bm{Q}_\mathrm{max} \approx \left( 0, \pi \right)$, which indicates enhanced AFM fluctuations between unit cells along the $c$-axis. 
Comparison with experimental results\cite{Kobayashi2017, Kobayashi2020_chem-press_NMR} suggests a similar temperature dependence of magnetic fluctuations. 
Both the experimental observations and the present study demonstrate that lowering the temperature induces a change in magnetic behavior from the AFM to the SDW-like fluctuations. 

Future studies should investigate the role of inter-molecular Coulomb interactions and their effects on physical properties such as the charge disproportionation and charge fluctuations in $\lambda$-(BETS)$_{2}$GaCl$_{4}$. 
Indeed, experimental results remain controversial regarding charge fluctuations, with some reports observing no enhancement~\cite{Kobayashi2020_no_increase_in_charge-fluctuation_NMR}, whereas others suggest the presence of charge disproportionation~\cite{Hiraki2010, Iakutkina2021}, which is related to charge fluctuations. 
It has also been pointed out that off-site Coulomb interactions play a role in inducing magnetic properties such as ferrimagnetism~\cite{Aizawa2020}. 
Thus, clarifying the effects of off-site Coulomb interactions remains as an important topic for future research. 

\section*{Acknowledgments}

The author is grateful to T. Kobayashi, S. Fukuoka, and K. Yoshimi for valuable discussions. 
The author would also like to acknowledge K. Kuroki for hosting him as a visiting researcher and Osaka University for providing research facilities. 
This work was partially supported by the Japan Society for the Promotion of Science~(JSPS) KAKENHI Grant Number 25K07236.


\begin{thebibliography}{99} 

\bibitem{H_Kobayashi1993}
H. Kobayashi, T. Udagawa, H. Tomita, T. Naito, K. Bun, T. Naito, and A. Kobayashi, 
Chem. Lett. \textbf{22}, 1559 (1993). 

\bibitem{A_Kobayashi1993}
A. Kobayashi, T. Udagawa, H. Tomita, T. Naito, and H. Kobayashi, 
Chem. Lett. \textbf{22}, 2179 (1993). 

\bibitem{Tanatar1999}
M. A. Tanatar, T. Ishiguro, H. Tanaka, A. Kobayashi, and H. Kobayashi, 
J. Supercond. \textbf{12}, 511 (1999). 

\bibitem{Yasuzuka2014}
S. Yasuzuka, S. Uji, T. Terashima, S. Tsuchiya, K. Sugii, B. Zhou, A. Kobayashi, and H. Kobayashi, 
J. Phys. Soc. Jpn. \textbf{83}, 013705 (2014). 

\bibitem{Imajo2016}
S. Imajo, N. Kanda, S. Yamashita, H. Akutsu, Y. Nakazawa, H. Kumagai, T. Kobayashi, and A. Kawamoto, 
J. Phys. Soc. Jpn. \textbf{85}, 043705 (2016). 

\bibitem{Imajo2019}
S. Imajo, S. Yamashita, H. Akutsu, H. Kumagai, T. Kobayashi, A. Kawamoto, and Y. Nakazawa, 
J. Phys. Soc. Jpn. \textbf{88}, 023702 (2019). 

\bibitem{Kobayashi2020_d-wave_NMR}
T. Kobayashi, H. Taniguchi, A. Ohnuma, and A. Kawamoto, 
Phys. Rev. B \textbf{102}, 121106(R) (2020). 

\bibitem{Dita2021}
D. P. Sari, R. Asih, K. Hiraki, T. Nakano, Y. Nozue, Y. Ishii, A. D. Hillier, and I. Watanabe, 
Phys. Rev. B \textbf{104}, 224506 (2021). 

\bibitem{Aizawa2018}
H. Aizawa, T. Koretsune, K. Kuroki, and H. Seo, 
J. Phys. Soc. Jpn. \textbf{87}, 093701 (2018). 

\bibitem{Tanatar2002}
M. A. Tanatar, T. Ishiguro, H. Tanaka, and H. Kobayashi, 
Phys. Rev. B \textbf{66}, 134503 (2002). 

\bibitem{Coniglio2011}
W. A. Coniglio, L. E. Winter, K. Cho, and C. C. Agosta, B. Fravel, and L. K. Montgomery, 
Phys. Rev. B \textbf{83}, 224507 (2011). 

\bibitem{Uji2015}
S. Uji, K. Kodama, K. Sugii, T. Terashima, T. Yamaguchi, N. Kurita, S. Tsuchiya, T. Konoike, M. Kimata, A. Kobayashi, B. Zhou, and H. Kobayashi, 
J. Phys. Soc. Jpn. \textbf{84}, 104709 (2015). 

\bibitem{Imajo2021}
S. Imajo, T. Kobayashi, A. Kawamoto, K. Kindo, and Y. Nakazawa, 
Phys. Rev. B \textbf{103}, L220501 (2021). 

\bibitem{Takagi2003_77Se_NMR}
S. Takagi, D. Maruta, H. Sasaki, H. Uozaki, H. Tsuchiya, Y. Abe, Y. Ishizaki, E. Negishi, H. Matsui, S. Endo, and N. Toyota, 
J. Phys. Soc. Jpn. \textbf{72}, 483 (2003). 

\bibitem{Takagi2003_1H_NMR}
S. Takagi, D. Maruta, H. Uozaki, H. Tsuchiya, Y. Abe, Y. Ishizaki, E. Negishi, H. Matsui, S. Endo, and N. Toyota, 
J. Phys. Soc. Jpn. \textbf{72}, 3259 (2003). 

\bibitem{Kobayashi2017}
T. Kobayashi and A. Kawamoto, 
Phys. Rev. B \textbf{96}, 125115 (2017). 

\bibitem{Kobayashi2020_no_increase_in_charge-fluctuation_NMR}
T. Kobayashi, K. Tsuji, A. Ohnuma, and A. Kawamoto, 
Phys. Rev. B \textbf{102}, 235131 (2020). 

\bibitem{Kobayashi2020_chem-press_NMR}
T. Kobayashi, T. Ishikawa, A. Ohnuma, M. Sawada, N. Matsunaga, H. Uehara, and A. Kawamoto
Phys. Rev. Res. \textbf{2}, 023075 (2020). 

\bibitem{Sawada2021}
M. Sawada, A. Kawamoto, and T. Kobayashi
Phys. Rev. B \textbf{103}, 045112 (2021). 

\bibitem{Seo1997}
H. Seo and H. Fukuyama, 
J. Phys. Soc. Jpn. \textbf{66}, 3352 (1997). 

\bibitem{Vilk1997}
Y. M. Vilk and A. -M. S. Tremblay, 
J. Phys. I France \textbf{7}, 1309 (1997). 

\bibitem{Otsuki2012}
J. Otsuki, 
Phys. Rev. B \textbf{85}, 104513 (2012). 

\bibitem{Arya2015}
S. Arya, P. V. Sriluckshmy, S. R. Hassan, and A.-M. S. Tremblay, 
Phys. Rev. B \textbf{92}, 045111 (2015). 

\bibitem{Ogura2015}
D. Ogura and K. Kuroki, 
Phys. Rev. B \textbf{92}, 144511 (2015). 

\bibitem{Aizawa2015}
H. Aizawa, K. Kuroki, and J. Yamada, 
Phys. Rev. B \textbf{92}, 155108 (2015). 

\bibitem{Aizawa2018PRB}
H. Aizawa and K. Kuroki, 
Phys. Rev. B \textbf{97}, 104507 (2018). 

\bibitem{Aizawa2018JPCS}
H. Aizawa, 
J. Phys.: Conf. Ser. \textbf{969}, 012095 (2018). 

\bibitem{Miyahara2013} 
H. Miyahara, R. Arita, and H. Ikeda, 
Phys. Rev. B \textbf{87}, 045113 (2013). 

\bibitem{Nakamura2009}
K. Nakamura, Y. Yoshimoto, T. Kosugi, R. Arita, and M. Imada, 
J. Phys. Soc. Jpn. \textbf{78}, 083710 (2009). 

\bibitem{Nakamura2016}
K. Nakamura, Y. Nohara, Y. Yoshimoto, and Y. Nomura, 
Phys. Rev. B \textbf{93}, 085124 (2016). 

\bibitem{Misawa2020}
T. Misawa, K. Yoshimi, and T. Tsumuraya, 
Phys. Rev. Research \textbf{2}, 032072(R) (2020). 

\bibitem{Kato2025}
T. Kato, H. Ma,K. Yoshimi, T. Misawa, S. Kumagai, Y. Iida, Y. Sasaki, M. Sawada, J. Gouchi, T. Kobayashi, H. Taniguchi, Y. Uwatoko, H. Sato, N. Matsunaga, A. Kawamoto, and K. Nomura, 
Phys. Rev. B \textbf{112}, 104513 (2025). 

\bibitem{Aryasetiawan2004}
F. Aryasetiawan, M. Imada, A. Georges, G. Kotliar, S. Biermann, and A. I. Lichtenstein, 
Phys. Rev. B \textbf{70}, 195104 (2004). 

\bibitem{Imada2010}
M. Imada and T. Miyake, 
J. Phys. Soc. Jpn. \textbf{79}, 112001 (2010). 

\bibitem{Hiraki2010}
K. Hiraki, M. Kitahara, T. Takahashi, H. Mayaffre, M. Horvati\'c, C. Berthier, S. Uji, H. Tanaka, B. Zhou, A. Kobayashi, and H. Kobayashi, 
J. Phys. Soc. Jpn. \textbf{79}, 074711 (2010). 

\bibitem{Iakutkina2021}
O. Iakutkina, E. Uykur, T. Kobayashi, A. Kawamoto, M. Dressel, and Y. Saito, 
Phys. Rev. B \textbf{104}, 045108 (2021). 

\bibitem{Aizawa2020}
H. Aizawa, 
J. Phys. Soc. Jpn. \textbf{89}, 114702 (2020). 

\end{thebibliography}
\end{document}